\documentclass[aps,twocolumn,nofootinbib,preprintnumbers,superscriptaddress]{revtex4}

\usepackage{color}
\usepackage{bbm}
\usepackage{tikz}

\newcommand{\nb}[1]{\color{blue}}

\newcommand{\hl}[1]{\color{magenta}}

\usepackage[
pagebackref=false,
colorlinks=true,
linkcolor=blue,
urlcolor=blue,
filecolor=black,
citecolor=red,
pdfstartview=FitV,
pdftitle={},
pdfauthor={},
pdfsubject={},
pdfkeywords={},
pdfpagemode=None,
bookmarksopen=true
]{hyperref}

\usepackage[normalem]{ulem}
\usepackage{amsmath, amssymb, bm}
\usepackage{enumerate}
\usepackage{amsfonts}
\usepackage{epsfig}
\usepackage{mathbbol}
\usepackage{mathrsfs}
\usepackage{braket}
\usepackage{subcaption}

\usepackage[utf8]{inputenc}
\DeclareUnicodeCharacter{0304}{\ensuremath{\bar{}}}
\usepackage{newtxtext,newtxmath}

\begin{document}

\title{
  Higher-order topological superconductor phases in a multilayer system
}

\author{Ryota Nakai}
\affiliation{Department of Physics, Kyushu University, Fukuoka, 819-0395, Japan}
\author{Kentaro Nomura}
\affiliation{Department of Physics, Kyushu University, Fukuoka, 819-0395, Japan}
\date{\today}

\begin{abstract}
Higher-order topological phases are gapped phases of matter that host gapless corner or hinge modes. 
For the case of superconductors, corner or hinge modes are gapless Majorana modes or Majorana zero modes.
To construct 3d higher-order topological superconductors, we consider a topological-insulator/superconductor multilayer under in-plane Zeeman coupling.
We found three different types of higher-order topological superconductor phases, a second-order topological superconductor phase with Majorana hinge flat bands,
a second-order Dirac superconductor phase with surface Majorana cones and Majorana hinge arcs, and nodal-line superconductor phases with drumhead surface states and Majorana hinge arcs.
\end{abstract}

\maketitle
\tableofcontents

\section{Introduction}

Topological insulators (TI) and topological superconductors (TSC) are electronic systems that have gapped bulk spectrum and gapless boundary modes.
These systems have been initially classified by the Altland-Zirnbauer symmetry class by the presence/absence of time-reversal, particle-hole, and chiral symmetries \cite{PhysRevB.78.195125,doi:10.1063/1.3149495}.
Beyond this classification, topological semimetals \cite{Shuichi_Murakami_2007,PhysRevB.83.205101,PhysRevLett.107.127205,RevModPhys.90.015001} that have gapless bulk nodes, 
topological crystalline insulators \cite{PhysRevLett.106.106802,doi:10.1146/annurev-conmatphys-031214-014501} that are protected by crystalline symmetries, 
and their corresponding superconductors (SC) \cite{PhysRevLett.90.057002,PhysRevB.86.054504,PhysRevLett.113.046401,Schnyder_2015,PhysRevB.88.075142,PhysRevB.90.165114} have been widely studied.

Topological systems show quantized responses directly related to topological numbers of the electronic states in filled bands.
In two dimensions, quantum Hall, quantum anomalous Hall (QAH), quantum spin Hall (QSH), and TSC systems show the quantized (electric, spin, thermal) Hall conductivity \cite{PhysRevLett.61.2015,PhysRevLett.95.146802,PhysRevB.61.10267}.
In one dimension, electronic systems with inversion symmetry have quantized dipole moment 
\cite{PhysRevB.83.245132,PhysRevLett.62.2747,PhysRevB.47.1651,PhysRevB.78.195424}.
The concept of the dipole-moment-quantization has been extended to the multipole-quantization such as quadrupole and octupole moments \cite{doi:10.1126/science.aah6442,PhysRevB.96.245115}.
Electronic systems hosting quantized multipole moments are known as higher-order TIs \cite{doi:10.1126/science.aah6442,PhysRevB.96.245115,doi:10.1126/sciadv.aat0346,PhysRevB.99.245151,Xie2021}.
Multipoles are quantized in the presence of crystalline symmetry such as reflection, rotation, and inversion \cite{PhysRevB.97.205136}, and hence the higher-order TIs are topological crystalline insulators beyond the conventional classification of topological insulators.
In general, $n$th-order TIs host gapless modes in a $d-n$ dimensional subspace where $d$ is the material's dimensions.
Specifically, while 1st-order TIs are usual TIs, 2nd-order TIs have corner zero modes in two dimensions and gapless hinge modes in three dimensions.
Higher-order topological phases has been confirmed in photonic systems \cite{2018Natur.555..342S,2018Natur.555..346P,2018NatPh..14..925I}

Of particular interest are the higher-order topological superconductors (HOTSC)
\cite{PhysRevLett.119.246402,PhysRevLett.119.246401,PhysRevB.97.094508,PhysRevB.97.205134,PhysRevB.98.165144, PhysRevB.96.195109,PhysRevLett.121.186801,PhysRevB.99.195431,PhysRevB.97.205135,PhysRevB.100.205406, PhysRevB.99.125149,PhysRevB.99.020508,PhysRevResearch.1.032048,PhysRevB.100.235302,PhysRevB.100.020509, PhysRevLett.122.187001,PhysRevLett.123.167001,PhysRevLett.123.036802,PhysRevB.101.220506,PhysRevResearch.2.033495, PhysRevB.103.045424,PhysRevB.104.134508,PhysRevB.104.L180503,PhysRevB.105.014509,PhysRevB.105.155423,PhysRevB.108.094504,PhysRevB.107.094512,zhang2022bulkvortex,PhysRevB.103.L140502,PhysRevB.106.214510}
as they host localized Majorana zero modes, which can be utilized in quantum computation \cite{RevModPhys.80.1083}.
Specifically, 2d second-order topological superconductor (SOTSC) phases have been predicted in heterostructures such as 2d TI/$d$-wave or $s_{\pm}$-wave SC \cite{PhysRevLett.121.096803,PhysRevLett.121.186801,PhysRevB.98.245413},
second-order TI/SC \cite{PhysRevB.100.205406,PhysRevLett.121.196801}, $p\pm ip$/$d$-wave SCs \cite{PhysRevB.98.165144},
3d antiferromagnetic TI/SC \cite{PhysRevB.99.195431}, 2d TI/$s$-wave SC under an in-plane magnetic field \cite{PhysRevLett.123.156801,PhysRevLett.124.227001,Wu_2022,Lu_2022,Chen_2023},
Rashba-metal/$s+id$-wave SC \cite{PhysRevLett.122.236401,PhysRevLett.125.017001}, a $\pi$-junction of SC/QSH-bilayer/SC \cite{PhysRevB.102.195401} and so on.
However, not so much is known about the construction of 3d HOTSC phases by heterostructures.

For the construction of novel three-dimensional topological phases, one possible route is to make a multilayer structure of topological materials and other gapped materials.
This strategy has been adopted for creating Weyl semimetal (WSM) phases by multilayers of TI/trivial insulator and WSM/trivial insulator \cite{PhysRevLett.107.127205,PhysRevB.95.155101}.
The method has been applied to multilayers of SC/TI and SC/WSM to construct Weyl superconductor (WSC) phases \cite{PhysRevB.86.054504,PhysRevB.101.094510}, which have point nodes in the bulk superconducting states and Majorana Fermi arcs on the surface.
Recently, a $\pi$-junction through two Rashba-metal layers under an in-plane magnetic field has been predicted to be a 2d SOTSC \cite{PhysRevLett.122.126402}.
This system is similar to a part of the SC/TI multilayer, as both systems require a $\pi$-junction ($\pi$-phase difference) between neighboring SCs \cite{PhysRevB.86.054504}.
So it is natural to expect that a multilayer of SC and TI under an in-plane magnetic field shows 3d HOTSC phases.

\begin{figure}
  \centering
  \includegraphics[width=\columnwidth]{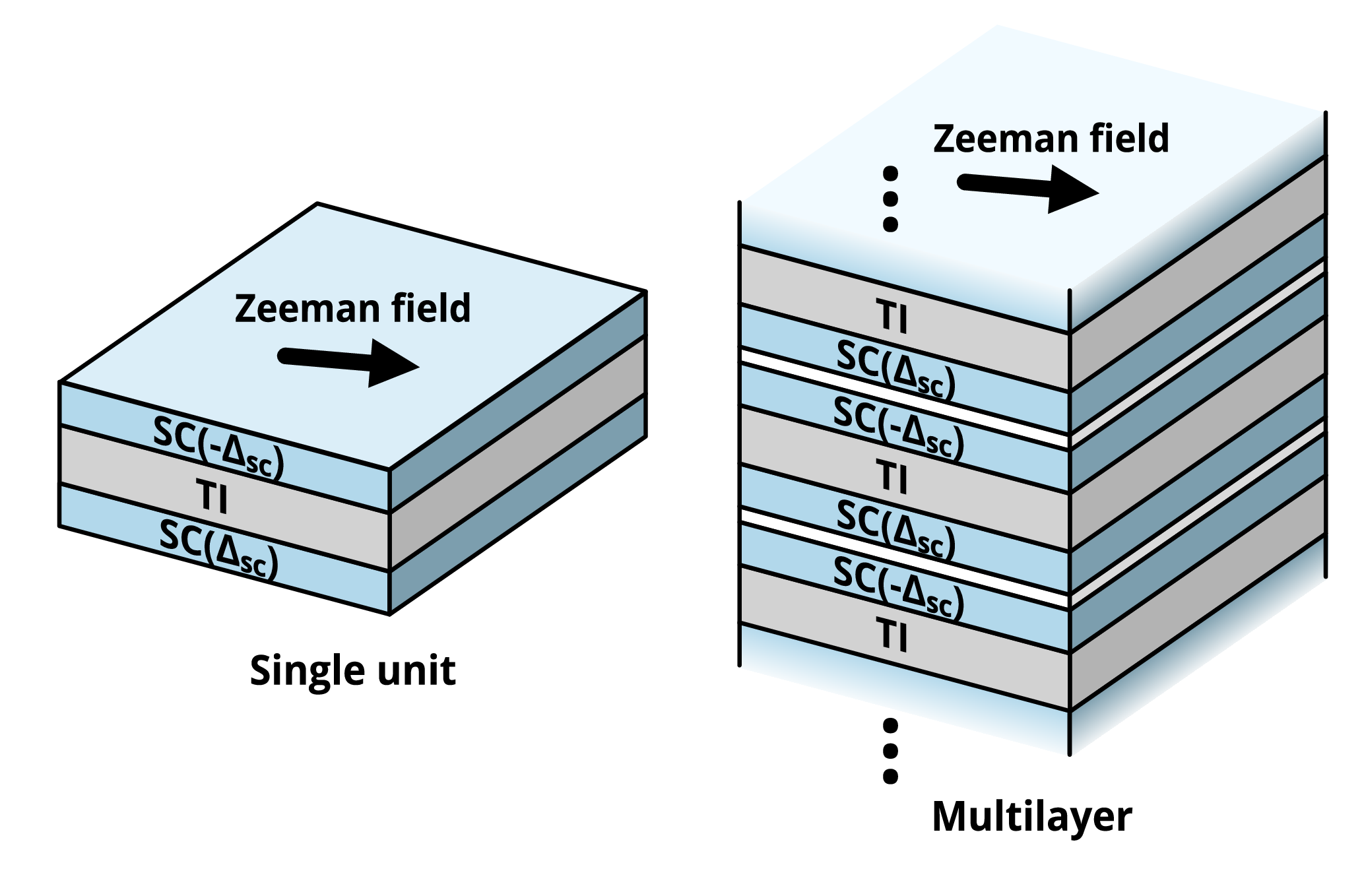}
  \caption{A single unit of superconductor(SC)/topological insulator(TI)/SC layers and a multilayer model of TI/SC under in-plane Zeeman coupling.
  In the multilayer model, each SC layer consists of two SC layers [SC($\Delta_\text{SC}$) and SC($-\Delta_\text{SC}$)] forming a $\pi$-junction separated by an insulator layer.
    \label{fig:multilayer}
  }
\end{figure}

In this work, we study a multilayer of TI and conventional SC forming $\pi$ junction under in-plane Zeeman coupling [Fig.~\ref{fig:multilayer} (right)].
The phase difference between neighboring SC layers can be tuned by a magnetic flux inserted through a Josephson junction whose endpoints are connected to the SC layers [not explicit in Fig.~\ref{fig:multilayer} (right), see \cite{PhysRevB.86.054504,PhysRevB.96.019901}].
Let us see what could happen under in-plane Zeeman coupling.
When the Zeeman field aligns out-of-plane direction, WSC phases appear between weak TSC phases whose sections (let's say normal to the $k_z$ axis) in the momentum space have nontrivial Chern numbers \cite{PhysRevB.86.054504}.
By flipping the Zeeman field, the sign of the Chern number is reversed.
With this mechanism, when the Zeeman field is tilted to the in-plane direction, weak TSC phases turn to nodal-line superconductor (NLSC) phases by closing a bulk gap at each $k_z$.
Similarly, the WSC phases also turn to NLSC phases, where nodes in a WSC phase are connected by nodal lines. 
In addition to NLSC phases, we found a 3d SOTSC phase and a second-order Dirac superconductor (SODSC) phase \cite{PhysRevB.100.020509}.
The 3d SOTSC is a bulk-gapped SC where the $k_z$ section is a 2d SOTSC.
The SODSC is a bulk-gapped SC where the $k_z$ section is either a 2d SOTSC or a trivial SC, and surface Majorana point nodes separate them.
The emergence of SOTSC phases in this multilayer system has been overlooked in previous works as these phases cannot be deduced from bulk and surface spectra.  
In addition, we found NLSC phases that appear next to the 3d SOTSC phase have both drumhead surface states and Majorana hinge states forming a flat band.

This paper is organized as follows.
We first study effective low-energy models of a TI/SC multilayer under in-plane Zeeman coupling in Sec.~\ref{sec:effective}.
We review the case with a single layer and thoroughly study the $\pi$-junction case as these results are used in the multilayer case.
Then we study the effective multilayer model.
To make the presence of Majorana hinge modes clear, we study numerically a lattice multilayer model in Sec.~\ref{sec:lattice}.
Similarly to the previous section, we start with a detailed study of a single-layer model and then move on to the multilayer case.
We finally conclude in Sec.~\ref{sec:conclusion}.

\section{Effective surface model}
\label{sec:effective}

First, we consider an effective model, in which only the surface Dirac fermions of the 3d TI layer are taken into account as an effective low-energy theory.
SCs are incorporated as an induced pair potential by the proximity effect.

\subsection{Single unit}
\label{sec:effective_single}

Let us discuss a single unit case, that is, a single TI layer sandwiched by two SCs [Fig.~\ref{fig:multilayer} (left)].
This system is similar to models studied in \cite{PhysRevLett.122.126402,Lu_2022,PhysRevB.83.220510},
but we will review in particular the SOTSC phase and the corner Majorana modes in detail aiming at extending to the multilayer system.
The effective Hamiltonian reads
\begin{align}
  &H=\sum_{\bm{k}_\perp}\left(c_{\bm{k}_\perp}^\dagger \mathcal{H}_0(\bm{k}_\perp)c_{\bm{k}_\perp}^{\ }
  +c_{\bm{k}_\perp}^\dagger\mathcal{H}_\Delta(\bm{k}_\perp)c_{-\bm{k}_\perp}^\dagger+\text{H.c.}\right), \notag\\
  &\mathcal{H}_0(\bm{k}_\perp)
  =v_F(\hat{z}\times\bm{\sigma})\cdot \bm{k}_\perp \rho^z+t_s\rho^x+\Delta_\text{Z}\bm{n}\cdot\bm{\sigma},\notag\\
  &\mathcal{H}_\Delta(\bm{k}_\perp)
  =i\Delta_\text{SC}\sigma^y(\cos\varphi/2+i\rho^z\sin\varphi/2),
\end{align}
where $c_{\bm{k}_\perp}=(\psi_{\bm{k}_\perp\text{top}\uparrow},\psi_{\bm{k}_\perp\text{top}\downarrow},\psi_{\bm{k}_\perp\text{bottom}\uparrow},\psi_{\bm{k}_\perp\text{bottom}\downarrow})$, 
$\bm{k}_\perp=(k_x,k_y)$ is the momentum along the TI surface, 
$\sigma$ and $\rho$ are the Pauli matrices for the spin and top/bottom-surface degrees of freedom, respectively,
$v_F$ is the Fermi velocity of the surface Dirac fermion, and $\hat{z}$ is the unit vector normal to the surfaces.
$\Delta_\text{SC}$ is the induced pair potential and $\varphi$ is the phase difference between the pair potential from the top and bottom SCs. 
$\Delta_\text{Z}$ is the magnitude of Zeeman coupling and $\bm{n}$ is the orientation of the Zeeman field.
The intra-layer tunneling amplitude between the top and bottom surfaces of the 3d TI is $t_s$.

In the Nambu space, the total Hamiltonian reads
\begin{align}
  &\mathcal{H}_\text{single}(\bm{k}_\perp)=v_F(-k_x\sigma^y\rho^z\tau^z+k_y\sigma^x\rho^z)+t_s\rho^x\tau^z\notag\\
  &\quad+\Delta_\text{Z}(n_x\sigma^x\tau^z+n_y\sigma^y+n_z\sigma^z\tau^z) \notag\\
  &\quad-\Delta_\text{SC}\sigma^y(\tau^y\cos\varphi/2+\rho^z\tau^x\sin\varphi/2),
  \label{eq:singlelayereffectivehamiltonian}
\end{align}
where $\tau$ is the Pauli matrix acting on the Nambu space.
This model is invariant under particle-hole conjugation by $\mathcal{P}=\tau^x\mathcal{K}$, where $\mathcal{K}$ is the complex conjugation operator.
Provided the phase difference is either 0 or $\pi$ and Zeeman coupling is absent, the model is invariant under time reversal by $\Theta=i\sigma^y\mathcal{K}$ when $\varphi=0$ and $\Theta=i\sigma^y\tau^z\mathcal{K}$ when $\varphi=\pi$ \cite{PhysRevB.83.220510}.
Regarding crystalline symmetry, the model without Zeeman coupling has mirror symmetry $M_x=\sigma^x\tau^z$ and $M_y=\sigma^y$ with respect to the plane normal to the $x$ and $y$ axes, respectively, and 4-fold rotational symmetry $U_4=e^{i\pi\sigma^z\tau^z/4}$ with respect to the $z$ axis.
Notice that Zeeman coupling can break mirror and rotational symmetries mentioned above.
Combining mirror symmetry $M_z=\sigma^z\rho^x\tau^x$ with respect to the plane normal to the $z$ axis, the total Hamiltonian is invariant under 3d inversion $I=M_xM_yM_z\propto\rho^x$.

Let us first discuss the phase diagram without Zeeman coupling ($\Delta_\text{Z}=0$).
Without the proximity effect ($\Delta_\text{SC}=0$), a thin TI layer is a quantum spin Hall (QSH) insulator \cite{PhysRevB.81.041307}.
When the proximity effect is turned on, the bulk phase depends on the phase difference $\varphi$.
The bulk spectrum has an energy gap $(t_s^2+\Delta_\text{SC}^2\pm 2t_s\Delta_\text{SC}\sin\varphi/2)^{1/2}$.
For the 0-junction ($\varphi=0$), the energy gap $(t_s^2+\Delta_\text{SC}^2)^{1/2}$ indicates that there is no phase transition between $t_s$-dominated region (QSH) and the $\Delta_\text{SC}$-dominated region, thus the resulting phase is a trivial SC phase.
From the edge perspective, the helical edge modes in QSH disappear by an energy gap induced by the proximity effect \cite{PhysRevB.79.161408}.
On the other hand, for the $\pi$-junction ($\varphi=\pi$), the energy gap is $t_s\pm \Delta_\text{SC}$, which indicates a topological phase transition at $|\Delta_\text{SC}|=|t_s|$.
The phase with $\Delta_\text{SC}/t_s>1$ is the helical TSC phase which is protected by time-reversal symmetry \cite{PhysRevB.83.220510}. 
When the phase difference is slightly changed from $\varphi=\pi$, the helical TSC phase disappears as time-reversal symmetry is broken.

\begin{figure}
  \centering
  \includegraphics[width=0.47\textwidth]{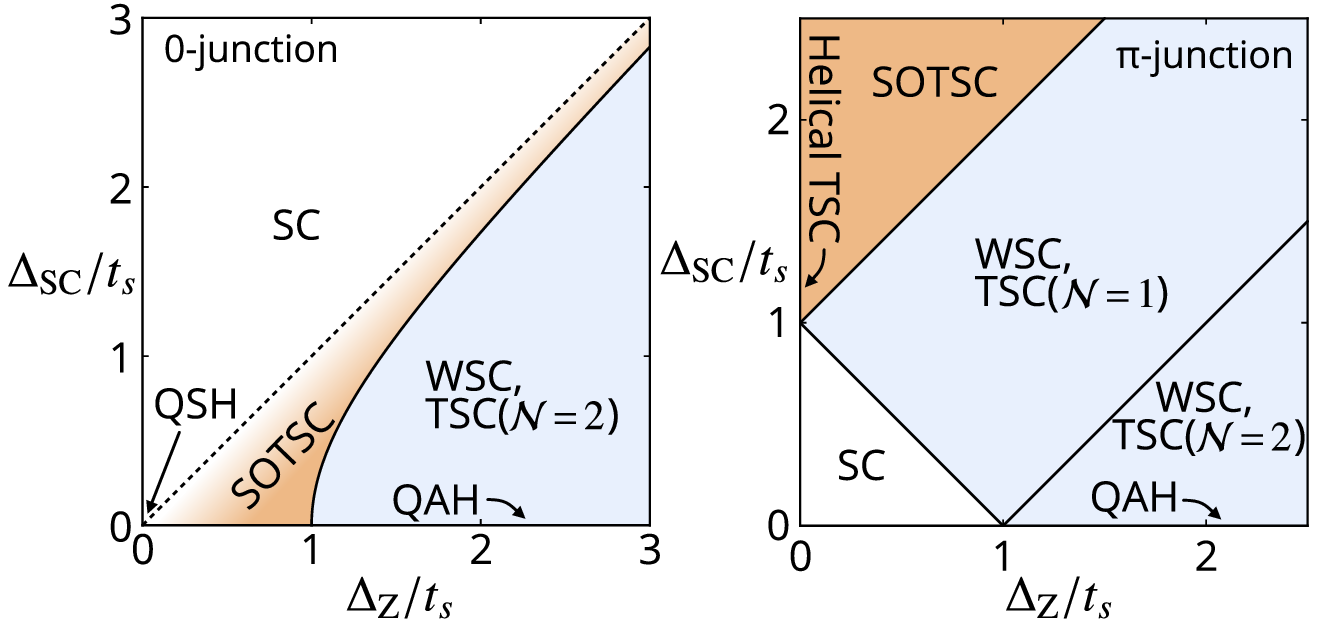}
  \caption{The phase diagrams of a single unit of SC/TI/SC Josephson junction, where the phase difference is 0 (left) and $\pi$ [right, see Fig.~\ref{fig:multilayer} (left)], are shown
  in $\Delta_\text{Z}$-$\Delta_\text{SC}$ space.
  Large Zeeman coupling regions ($\Delta_\text{Z}\gg\Delta_\text{SC}$) are dominated by topological superconductor (TSC) phases when an out-of-plane Zeeman field is applied and Weyl superconductor (WSC) phases when in-plane.
  A second-order topological superconductor (SOTSC) phase with two Majorana corner modes appears in the large pair-potential region ($\Delta_\text{Z}\ll\Delta_\text{SC}$) in a $\pi$-junction.
    \label{fig:singlephasediagram}
  }
\end{figure}

Including Zeeman coupling in the 0-junction,
the spectrum is given by
\begin{align}
 \epsilon^2=v_F^2(k_x^2+k_y^2\cos^2\theta)+\left(\Delta_\text{Z}\pm\sqrt{v_F^2k_y^2\sin^2\theta+t_s^2+\Delta_\text{SC}^2}\right)^2,
\end{align}
where $\bm{n}=(\sin\theta,0,\cos\theta)$.
All states are doubly degenerate.
Unless the Zeeman field is oriented in-plane ($\theta\neq\pi/2$), the energy gap closes at $\bm{k}_\perp=(0,0)$ when $\Delta_\text{Z}=\pm(t_s^2+\Delta_\text{SC}^2)^{1/2}$.
The phase diagram is shown in Fig.~\ref{fig:singlephasediagram} (left).
The Zeeman-coupling-dominated phase is a time-reversal-symmetry-broken TSC phase with the Bogoliubov-de Gennes Chern number $\mathcal{N}=\pm2$.
The Chern number is determined by the sign of the $z$ component of the Zeeman field ($\text{sign}[\Delta_\text{Z}n_z]$).
Hence, when the Zeeman field aligns the in-plane direction ($\theta=\pi/2$), the Zeeman-coupling-dominated phase is the topological phase transition point between TSC phases with $\mathcal{N}=2$ and $-2$.
This phase is a 2d WSC phase, which has point nodes in the bulk.
Notice that the energy gap of the helical edge mode opened by the proximity effect can be closed and reopened by Zeeman coupling.
When Zeeman coupling works at different magnitudes for different edges in a rectangular geometry, corner Majorana modes could appear between $\Delta_\text{SC}$-dominated and $\Delta_\text{Z}$-dominated edges \cite{PhysRevB.79.161408}.
This phase is a SOTSC phase with 4 corner Majorana modes \cite{PhysRevLett.124.227001}. 
However, the appearance of these corner Majorana modes depends on the orientation of the Zeeman field and does not accompany a bulk topological phase transition as this is solely an edge phenomenon.
This phase is not the focus of this study.

On the other hand, including Zeeman coupling in a $\pi$-junction, the eigenenergy is given by
\begin{align}
  \epsilon^2=v_F^2(k_x^2+k_y^2\cos^2\theta)+\left(\Delta_\text{Z}\pm\sqrt{v_F^2k_y^2\sin^2\theta+(t_s\pm\Delta_\text{SC})^2}\right)^2.
\end{align}
Unless in-plane ($\theta\neq 0$), there are topological phase transitions at $\Delta_\text{Z}=\pm t_s\pm\Delta_\text{SC}$ [Fig.~\ref{fig:singlephasediagram} (right)].
Between the TSC phase with $\mathcal{N}=2$ (continuously connected to the QAH phase) and the SC phase, a TSC phase with $\mathcal{N}=1$ appears, which is similar to Ref.~\cite{PhysRevB.82.184516}.
Similarly to the 0-junction case, when the Zeeman field aligns the in-plane direction, TSC phases become WSC phases.
The focus of this study is the SOTSC phase when $\Delta_\text{SC}$ is dominated.

Let us analyze this phase from the edge perspective following \cite{PhysRevB.83.220510}.
Without Zeeman coupling, the helical TSC phase has helical Majorana edge modes that are protected by time-reversal symmetry.
Consider an edge at $y=0$ where the bulk is in $y>0$.
The helical Majorana modes with $k_x=0$ is the solution of  $\mathcal{H}_\text{single}(0,-i\partial_y)\psi(y)=0$ by slightly changing the tunneling term as $t_s\to t_s-B\partial_y^2\,(B>0)$.
To satisfy the boundary condition $\psi(y=0)=0$ and $\psi(y\to\infty)=0$, the equation is solved with an ansatz wavefunction 
\begin{align}
  \psi(y)\propto(e^{-\eta_+ y}-e^{-\eta_- y})\phi,
\end{align}
where $\phi$ is an 8-component spinor.
Two solutions $\psi^\text{R/L}(y)$ are present provided $\Delta_\text{SC}>t_s$ and are given by $\eta_\pm=[v_F\pm(v_F^2+4B(t_s-\Delta_\text{SC}))^{1/2}]/2B$ with eigenspinors $\phi=\phi^\text{R/L}$, respectively.
Since $\phi^\text{R}$ ($\phi^\text{L}$) is the eigenvector of $-v_Fk_x\sigma^y\rho^z\tau^z$ with eigenvalues $v_Fk_x$ ($- v_Fk_x$), these correspond to the helical Majorana edge modes.

The helical Majorana modes are gapped by Zeeman coupling \cite{PhysRevB.97.205134,PhysRevLett.122.126402}.
Since $\langle\phi^\text{R}|\sigma^x\tau^z|\phi^\text{L}\rangle=i$, the $x$ component of Zeeman coupling opens a gap.
The gap is finite unless the Zeeman field is perpendicular to the edge, and the sign of the gap depends on the relative direction between the Zeeman field and the edge.
The orientation of the edge is defined counter-clockwise direction around the bulk.
In a rectangular geometry, the sign of the gap changes at two corners provided that the in-plane Zeeman field is not parallel to any of the edges.
These corners can be viewed as domain walls where the sign of the gap changes, and thus bind corner Majorana zero modes \cite{PhysRevLett.122.126402}.
Since the mass induced by in-plane Zeeman coupling is odd under 2-fold rotational symmetry $C_2$, the presence of the corner Majorana zero modes is guaranteed in rectangular geometry \cite{PhysRevB.97.205134,PhysRevB.97.205136,PhysRevB.97.205135}.
This phase is a 2d SOTSC phase with two Majorana corner modes.

\subsection{Multilayer}

Next, we extend the model in the previous section to a multilayer model by stacking TI and SC layers repeatedly [Fig.~\ref{fig:multilayer} (right)].
The case in which the Zeeman field aligns out-of-plane ($z$) direction has been studied in \cite{PhysRevB.86.054504}.
The low-energy effective Hamiltonian is given by
\begin{align}
  H=\sum_{i,j,\bm{k}_\perp}\Psi_{i\bm{k}_\perp}^\dagger \mathcal{H}_{ij}(\bm{k}_\perp)\Psi_{j\bm{k}_\perp},
\end{align}
where $i,j$ are the index of the TI layers, $\bm{k}_\perp=(k_x,k_y)$, and $\Psi_{i\bm{k}_\perp}=(c^{\ }_{i\bm{k}_\perp},c^\dagger_{i-\bm{k}_\perp})$.
The Hamiltonian of each TI layer is the same as the one in the previous section (\ref{eq:singlelayereffectivehamiltonian}),
and the tunneling terms couple neighboring layers as
\begin{align}
  \mathcal{H}_{ij}(\bm{k}_\perp)
  =
  &\mathcal{H}_\text{single}(\bm{k}_\perp)\delta_{ij}
  +\frac{t_d}{2}\left(\delta_{i,j+1}\rho^-+\delta_{i,j-1}\rho^+\right)\tau^z.
\end{align}
By Fourier transform in the $z$ direction, we obtain the Hamiltonian in the momentum space
\begin{align}
  \mathcal{H}(\bm{k})
  =
  \mathcal{H}_\text{single}(\bm{k}_\perp)
  &+t_d(\cos k_z\rho^x\tau^z+\sin k_z \rho^y\tau^z),
  \label{eq:multilayereffectivehamiltonian}
\end{align}
where $\bm{k}=(k_x,k_y,k_z)$.
This model has the same symmetries as the single-unit model, that is, particle-hole and inversion symmetries,
while time-reversal, 4-fold rotational, and mirror symmetries are broken by Zeeman coupling.

In the following, we focus on the $\pi$ phase difference and the $x$ component of the Zeeman field.
Comparing (\ref{eq:singlelayereffectivehamiltonian}) and (\ref{eq:multilayereffectivehamiltonian}),
the Hamiltonian of the multilayer model at $k_z=0$ and $\pi$ is the same as that of the single layer by replacing intra-layer coupling by $t_s\to t_s+t_d$ and $t_s-t_d$, respectively.
So, we can deduce the phase diagram of the multilayer model from the single-layer results.
In the following, we assume $t_s>t_d>0$, but the other case can be discussed in the same way.

\begin{figure}
  \centering
  \includegraphics[width=\columnwidth]{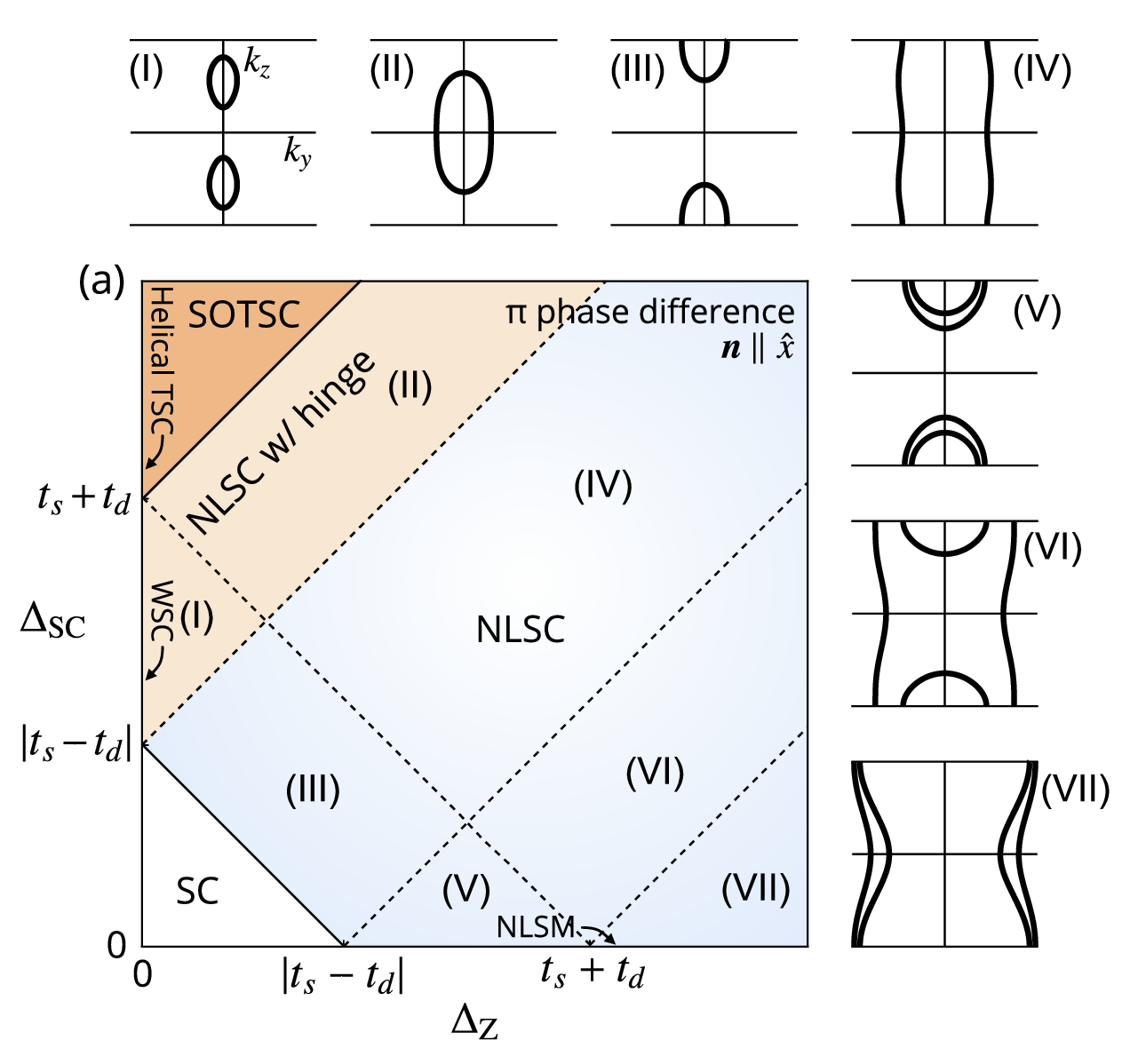}
  \includegraphics[width=\columnwidth]{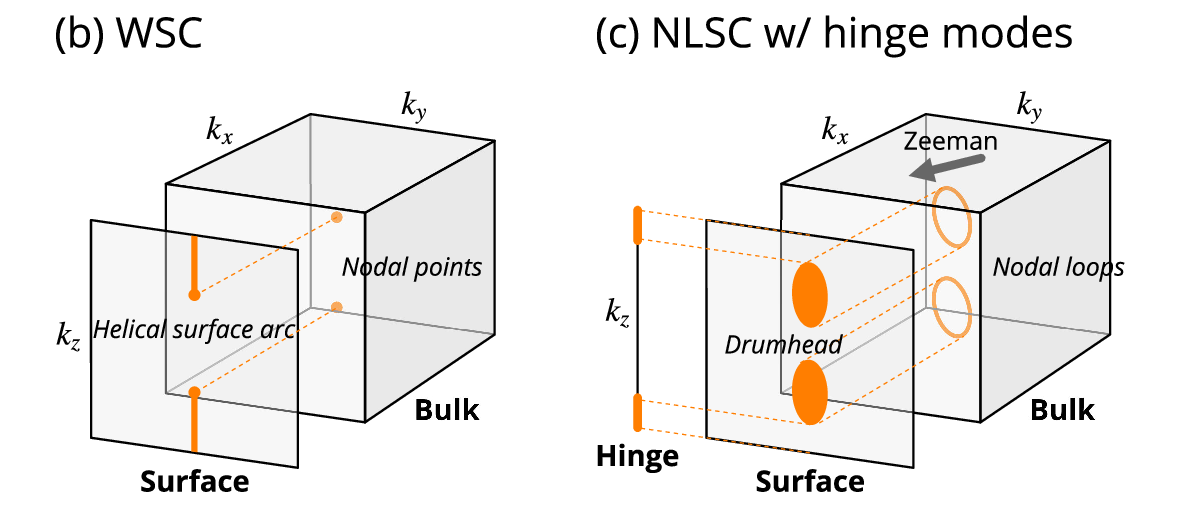}
  \caption{(a) The phase diagram of the effective multilayer model [Fig.~\ref{fig:multilayer} (right)] with $\pi$ phase difference in $\Delta_Z$-$\Delta_\text{SC}$ space,
  and the shape of line nodes in the bulk in the $k_y$-$k_z$ plane are shown.
  Large pair potential favors a three-dimensional second-order topological superconductor (SOTSC, the orange area) phase, while large Zeeman coupling favors nodal-line superconductor (NLSC, blue areas) phases.
  Between them, NLSC phases with drumhead surface modes and hinge modes (light orange regions) appear.
  The position of zero modes in the bulk, surface, and hinge in the momentum space is shown without (b) and with (c) in-plane Zeeman coupling, which correspond to Weyl superconductor (WSC) and NLSC phases in phase (I), respectively.
    \label{fig:effective_phasediagram}
  }
\end{figure}

Without Zeeman coupling, the phase with $\Delta_\text{SC}>t_s+t_d$ has helical Majorana surface modes in the entire $k_z$ space ($k_z\in[-\pi,\pi]$).
This phase is a 3d helical TSC phase, each $k_z$ section of which is a 2d helical TSC.
On the other hand, in the phase with $\Delta_\text{SC}\in[t_s-t_d,t_s+t_d]$, helical Majorana surface modes are present at $k_z=\pi$ but absent at $k_z=0$.
This indicates a helical Majorana Fermi arc on the surface and Weyl nodes in the bulk.
This is a 3d WSC phase [Fig.~\ref{fig:effective_phasediagram} (b)].

Including Zeeman coupling along $x$ direction, the helical TSC phase becomes a 3d SOTSC phase as an energy gap opens in the helical Majorana surface modes.
Majorana hinge modes form a flat band at $\epsilon=0$.
What is interesting in this model is the appearance of NLSC phases, which has no correspondence in the single-unit model [Fig.~\ref{fig:effective_phasediagram} (c)].
By increasing Zeeman coupling, the 3d SOTSC phase turns to an NLSC phase as the 2d SOTSC phase at the $k_z=0$ section turn to a 2d WSC phase [see Fig.~\ref{fig:singlephasediagram} (right)]. 
Let us check the bulk energy spectrum.
Here we consider the spectrum at $k_x=0$
\begin{align}
  &\epsilon(k_x=0,k_y,k_z)= \notag\\
  &\pm\Delta_\text{Z}\pm\left[(v_Fk_y)^2+\left(\Delta_\text{SC}\pm\sqrt{t_s^2+t_d^2\pm2t_st_d\cos k_z}\right)^2\right]^{1/2}.
\end{align}
Nodal line(s) appears in the $k_y$-$k_z$ plane.
The condition of the appearance of the bulk nodes is $|\Delta_\text{Z}|>|\Delta_\text{SC}\pm(t_s^2+t_d^2\pm 2t_st_d\cos k_z)^{1/2}|$.
Nodal lines are present in the region where this condition is satisfied.
The phase diagram on $\Delta_\text{SC}$-$\Delta_\text{Z}$ space is shown in Fig.~\ref{fig:effective_phasediagram} (a).
Along the dashed lines, the geometrical feature of the nodal-line shape changes.
Notice that as the orientation of the Zeeman field deviates from the in-plane direction, nodal lines are gapped out except at $\bm{k}_\perp=(0,0)$ and becomes a 3d WSC phase \cite{PhysRevB.86.054504}. 
In the limit of $\Delta_\text{SC}=0$, two nodal lines close to each other fuse and become a nodal line semimetal (NLSM) [(V) and (VII) phases in Fig.~\ref{fig:effective_phasediagram} (a)]. 

Specifically, in phase (I) in Fig.~\ref{fig:effective_phasediagram} (a), 
Zeeman coupling turns the bulk nodes into bulk nodal lines [Fig.~\ref{fig:effective_phasediagram} (b) into Fig.~\ref{fig:effective_phasediagram} (c)].
The nodal lines accompany drumhead states inside loops that are the projection of the bulk nodal lines onto the surface.
These drumhead states are bound to the surface.
Now the helical Majorana Fermi arc disappears as Zeeman coupling opens a gap.
Instead, zero modes appear at the hinge.
The Majorana hinge modes form a flat band in part of the Brillouin zone along $k_z$, whose endpoints are at the edges of the drumhead states [Fig.~\ref{fig:effective_phasediagram} (c)].
An explicit form of the wavefunction of the Majorana hinge modes can be obtained at $k_z=0$ and $\pi$ in the same manner as the single-layer case.
A similar situation happens also in the (II) phase in Fig.~\ref{fig:effective_phasediagram} (a). 
The number of Majorana hinge modes in (I) and (II) changes continuously between the SOTSC phase and the NLSC phases (III) and (IV) without Majorana hinge modes.
(I) and (II) phases are unique to a 3d multilayer system in the sense that these phases shrink in the limit of vanishing inter-layer coupling $t_d\to 0$, and the phase at each $k_z$ section is not uniform.
The origin of the Majorana hinge zero modes is the same as in the single-unit model as this multilayer model respects the same symmetries.

\section{Lattice model}
\label{sec:lattice}

In this section, we consider a lattice model to observe the corner Majorana modes.
In this model, the proximity effect is incorporated as an induced pair potential on the top and bottom layers of the TI lattice.

\subsection{Single unit}

First, we consider a single unit of an SC/TI/SC layer.
Consider the lattice model of the Bi$_2$Se$_3$-type TI given by \cite{PhysRevB.82.045122,Li2010}
\begin{align}
  \mathcal{H}_\text{TI}(\bm{k})=&
  \alpha(\sin k_x\sigma^x\lambda^x+\sin k_y\sigma^y\lambda^x+\sin k_z\lambda^y) \notag\\
  &+[m+t(3-\cos k_x-\cos k_y-\cos k_z)]\lambda^z,
  \label{eq:bisemodel}
\end{align}
where $\sigma$ and $\lambda$ are Pauli matrices for the spin and orbital degrees of freedom,
$\alpha$ is the amplitude of spin-orbit coupling.
By inverse Fourier transform with respect to $k_z$, we obtain the Hamiltonian in a slab geometry $\mathcal{H}_\text{TI}(\bm{k}_\perp,z,z')$.
This model shows strong TI phases when $m/t\in(-6,-4)$ and $(-2,0)$,
a weak TI phase when $m/t\in(-4,-2)$, and a trivial insulator phase otherwise.
For a thin film, the 2d bulk states show a QSH phase when $m/t\in(-5,-1)$ and a trivial insulator phase otherwise.
Within the QSH phase, there is a phase transition point at $m/t=-3$ where the position of the helical edge modes changes from $k=0$ to $\pi$.
We also consider Zeeman coupling and the proximity effect
\begin{align}
  &\mathcal{H}_\text{Z}(\bm{k}_\perp,z,z)=
  \Delta_\text{Z}(\bm{n}\cdot\bm{\sigma}), \\
  &\mathcal{H}_\Delta(\bm{k}_\perp,z,z)
  =
  \delta_{z,z_\text{top}}\Delta_\text{SC}(-\cos\varphi/2 \sigma^y\tau^y-\sin \varphi/2\sigma^y\tau^x) \notag\\
  &+\delta_{z,z_\text{bottom}}\Delta_\text{SC}(-\cos\varphi/2\sigma^y\tau^y+\sin \varphi/2\sigma^y\tau^x),
\end{align} 
where $z_\text{top}$ and $z_\text{bottom}$ are the position of the top and bottom surfaces of the TI lattice
and $\tau$ is the Pauli matrix on the Nambu space.
Here, we consider only a $\pi$-junction ($\varphi=\pi$).
This model has the same symmetry as the effective model in the previous section.

\begin{figure}
  \centering
  \includegraphics[width=\columnwidth]{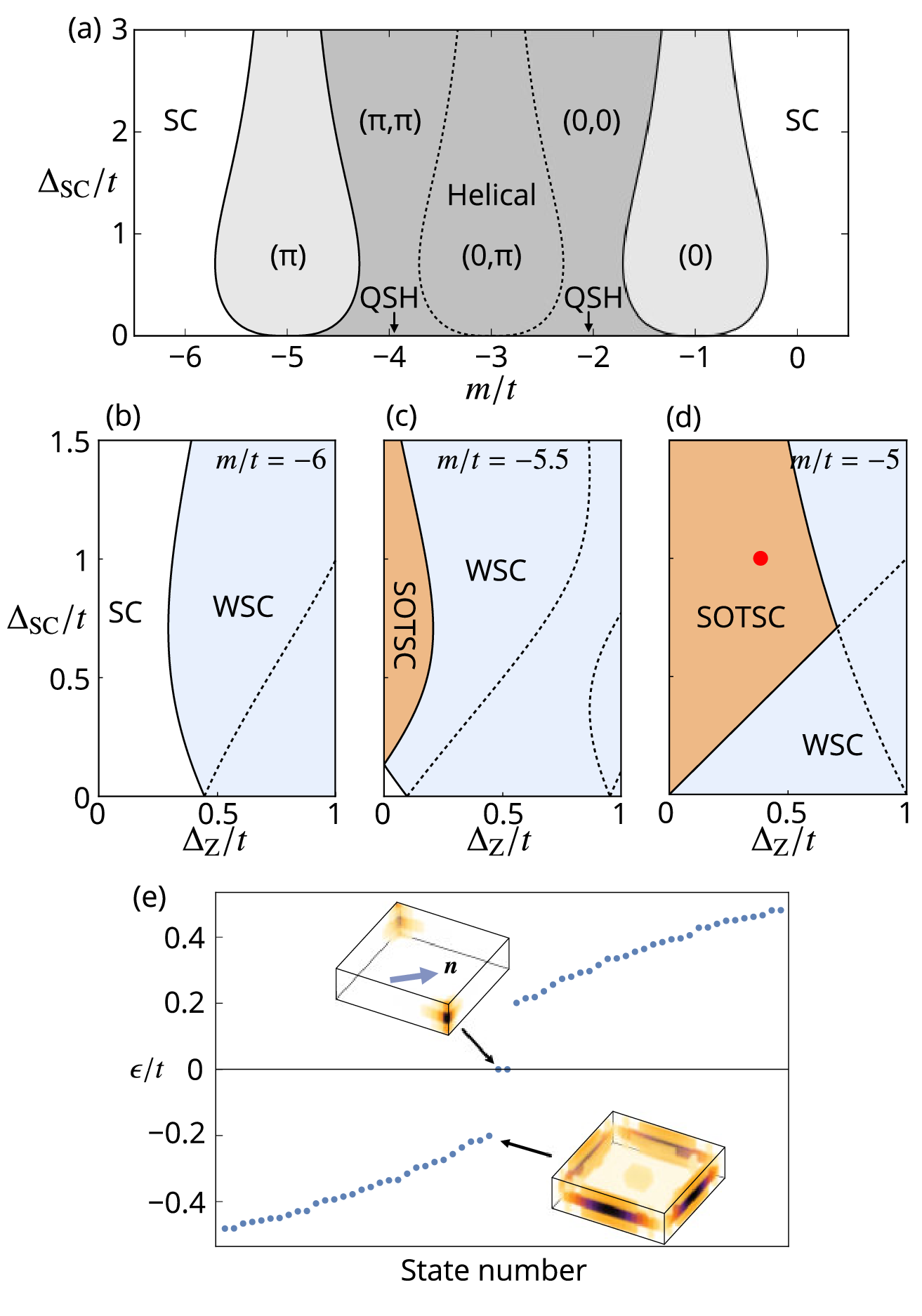}
  \caption{(a) The phase diagram of a $\pi$-junction through a 3-layer TI on $m$-$\Delta_\text{SC}$ space without Zeeman coupling is shown. Shaded regions are helical topological superconductor (TSC) phases, and the number and the position of helical edge Majorana modes are indicated.
  The phase diagrams with Zeeman coupling are shown on $\Delta_\text{Z}$-$\Delta_\text{SC}$ space at (b) $m/t=-6$, (c) $m/t=-5.5$, and (d) $m/t=-5$, respectively.
  Zeeman coupling turns helical TSC phases into second-order topological superconductor (SOTSC) phases.
  Even larger Zeeman coupling turns them into Weyl superconductor (WSC) phases.
  (e) The energy spectrum at $m/t=-5, \Delta_\text{SC}/t=1, \Delta_\text{Z}/t=0.4$ [red point in (d)] and the wavefunction of the corner Majorana modes (inset).
  \label{fig:latticesingle}}
\end{figure}

Without Zeeman coupling, helical TSC phases appear near the phase transition point at $m/t=-5$ and $-1$ [Fig.~\ref{fig:latticesingle} (a)].
The phase diagram is determined for a lattice model with 3 layers. 
The helical TSC phases near $m/t=-1$ and $-5$ have helical Majorana edge modes at $k=0$ or $\pi$, respectively.
Between two helical TSC phases, there are 3 different SC phases, each has 2 sets of helical Majorana modes.
The momenta at which the helical Majorana modes appear are shown in Fig.~\ref{fig:latticesingle} (a).

Zeeman coupling induces a 2d SOTSC phase close to the helical TSC phases, where 2 or 4 corner Majorana zero modes appear.
The number of the Majorana corner modes is equal to that of the helical Majorana edge modes without Zeeman coupling.
The energy spectrum of $20\times 20$ sites with thickness 3 sites is shown in Fig.~\ref{fig:latticesingle} (e), where we used parameters $\alpha/t=1, m/t=-5, \Delta_\text{SC}/t=1, \Delta_\text{Z}/t=0.4$, and $\bm{n}=(1/\sqrt{2},1/\sqrt{2},0)$.
With these parameters, the bulk energy gap is $\sim 0.5t$, and Zeeman coupling opens a gap in the edge modes leaving zero-energy Majorana corner modes.
In this system, the edge gap is opened by Zeeman coupling perpendicular to the edge (e.g., the $y$ component on the edge along $x$).
As a result, Majorana zero modes appear at corners where the sign of the $z$ component of $\bm{n}\times\bm{e}$ ($\bm{e}$ is the unit vector parallel to the edge) changes.
The position of the Majorana corner modes depends on the details of the model and is different from the effective low-energy theory in the previous section and \cite{PhysRevLett.122.126402}. 
Notice that if we use another model of TI with $\cos k_z\sigma^z\lambda^x$ in place of $\cos k_z\tau^y$ in (\ref{eq:bisemodel}) (e.g., see \cite{PhysRevB.82.045122}), an energy gap of the helical Majorana edge mode is opened by the orbital-dependent Zeeman coupling and the gap size is proportional to its edge-parallel component (e.g., for an edge along $x$, $\sigma^x\lambda^z\tau^z$).
In this case, the position of the corner Majorana modes is consistent with the effective model.

The phase diagram on $\Delta_\text{Z}$-$\Delta_\text{SC}$ space is shown for $m/t=-6,-5.5$, and $-5$ [Fig.~\ref{fig:latticesingle} (b), (c), and (d), respectively].
Within WSC phases, dashed lines indicate phase-transition lines where the number and/or the position of the bulk Weyl nodes changes.
When the phase without Zeeman coupling is a helical TSC (SC) phase, the gapped phase induced by Zeeman coupling is the SOTSC (SC) phase.
The phase diagram is symmetric with respect to $m/t=-3$.

\subsection{Multilayer}

Then we consider a multilayer model of the lattice TI and SC with alternating phases on top and bottom surfaces:
\begin{align}
  H=\sum_{i,j,\bm{k}_\perp,z,z'}\Psi_{i\bm{k}_\perp z}^\dagger \mathcal{H}_{ij}(\bm{k}_\perp,z,z')\Psi_{j\bm{k}_\perp z'}.
\end{align}
Here 
\begin{align}
  &\mathcal{H}_{ij}(\bm{k}_\perp,z,z')
  =\notag\\
  &\left[\mathcal{H}_\text{TI}(\bm{k}_\perp,z,z')+\mathcal{H}_\text{Z}(\bm{k}_\perp,z,z')+\mathcal{H}_\Delta(\bm{k}_\perp,z,z')\right]\delta_{ij} \notag\\
  &+\frac{t_d}{2}\left(\delta_{i,j+1}\delta_{z,z_\text{bottom}}\delta_{z',z_\text{top}}+\delta_{i,j-1}\delta_{z,z_\text{top}}\delta_{z',z_\text{bottom}}\right)\lambda^z\tau^z,
  \label{eq:hamiltonian_latticemultilayer}
\end{align}
where $i,j$ are the TI layer index and $z$ is the coordinate of each layer along the stacking direction.
We assume that the inter-layer coupling depends on the orbital ($\propto \lambda^z$).
If we employ an orbital-independent inter-layer coupling ($\propto \lambda^0$), the phase diagram of the multilayer model is the same as that of the single-layer model,
that is, the inter-layer coupling does not work at all and each layer is independent.
Notice that $t$ in (\ref{eq:bisemodel}) is a parameter of the Hamiltonian and is independent of the thickness of the TI layer, while $t_s$ in the effective model is the tunneling amplitude of the top and bottom surfaces and thus is dependent on the thickness.

\begin{figure}
  \centering
  \includegraphics[width=0.9\columnwidth]{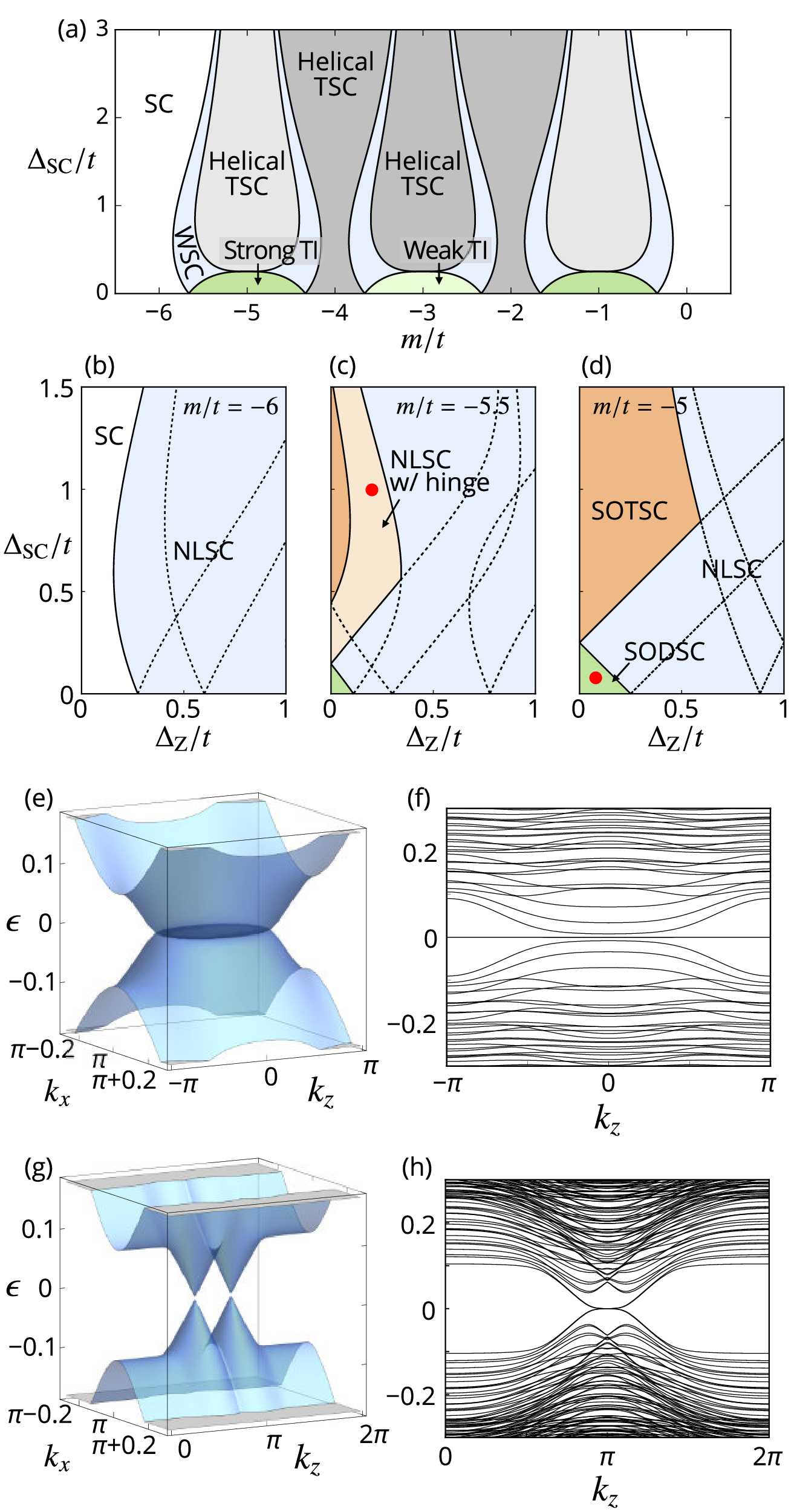}
  \caption{(a) The phase diagram of a multilayer lattice model on $m$-$\Delta_\text{SC}$ space without Zeeman coupling is shown.
  Weyl superconductor (WSC) phases intervene between SC and helical topological superconductor (TSC) phases or between helical TSC phases with different number/position of helical surface Majorana modes.
  Strong and weak topological insulator (TI) phases survive in the presence of a weak proximity effect.
  The phase diagrams with Zeeman coupling on $\Delta_\text{Z}$-$\Delta_\text{SC}$ space are shown at (b) $m/t=-6$, (c) $m/t=-5.5$, and (d) $m/t=-5$, respectively.
  Zeeman coupling turns helical TSC phases into second-order topological superconductor (SOTSC) phases, WSC phases into nodal-line superconductor (NLSC) phases with drumhead surface modes and hinge modes, and strong TI phases into second-order Dirac superconductor (SODSC) phases.
  The energy spectra of a NLSC with drumhead and hinge modes [$m/t=-5.5$, $\Delta_\text{Z}/t=0.2$, $\Delta_\text{SC}/t=1$ and $\bm{n}=(1/\sqrt{2},1/\sqrt{2},0)$ indicated by the red point in (c)] on a slab geometry with the thickness $L_y=200$ (e) and on geometry in Fig.~\ref{fig:multilayer} (right) with the size $(L_x,L_y)=(40,40)$ (f) are shown.
  (g) and (h) are the energy spectra of an SODSC [$m/t=-5$, $\Delta_\text{Z}/t=0.08$, and $\Delta_\text{SC}/t=0.08$ indicated by the red point in (d)] on the same geometry as (e) and (f), but the size of (h) is $(L_x,L_y)=(60,60)$.
  \label{fig:latticemulti}}
\end{figure}

Fig.~\ref{fig:latticemulti} (a) is the phase diagram without Zeeman coupling. 
Here we use $t_d/t=0.5$.
The bulk phase without the proximity effect ($\Delta_\text{SC}=\Delta_\text{Z}=0$) is a strong TI phase near $m/t=-5$ and $-1$ and a weak TI phase near $m/t=-3$.
There appear WSC phases between helical TSC and trivial SC phases or between helical TSC phases [Fig.~\ref{fig:latticemulti} (a)].
The Weyl nodes projected onto a surface are endpoints of a Fermi arc of helical surface Majorana modes [Fig.~\ref{fig:effective_phasediagram} (b)].
Any two helical TSC phases separated by a WSC phase have a different number or position of the helical Majorana modes.

By including in-plane Zeeman coupling, the helical TSC phase turns to a 3d SOTSC phase and WSC phases to NLSC phases [Fig.~\ref{fig:latticemulti} (b), (c), and (d)]. 
The Majorana hinge modes appear in the whole $k_z$ Brillouin zone in the SOTSC phase and 
and within a part of the Brillouin zone in some NLSC phases.
Phases with strong Zeeman coupling are mostly NLSC phases without hinge modes, which is consistent with the effective model.
Dashed lines in NLSC phases represent phase transitions where the number or the shape of the nodal line changes.
In an NLSC phase [the red point in Fig.~\ref{fig:latticemulti} (c)], the energy spectrum is shown for two geometries: a slab geometry with the open boundary condition (OBC) in the $y$ direction and the periodic boundary condition (PBC) in the other directions [Fig.~\ref{fig:latticemulti} (e)],
and in a geometry in Fig.~\ref{fig:multilayer} (right) under the OBC in both $x$ and $y$ directions while the PBC in the $z$ direction [Fig.~\ref{fig:latticemulti} (f)].
The drumhead states can be observed in the slab geometry while the Majorana hinge modes forming a $\epsilon=0$ flat band can be seen in the geometry in Fig.~\ref{fig:multilayer} (right).
The hinge modes are terminated at the edge of the drumhead states as schematically shown in Fig.~\ref{fig:effective_phasediagram} (c).

The strong and weak TI phases around $m/t=-5,-3$ and $-1$ retain surface Dirac cone(s) even in the presence of the weak proximity effect [green regions in Fig.~\ref{fig:latticemulti} (a)].
Small Zeeman coupling split the surface Dirac cone(s) into two (four) surface Majorana cones [Fig.~\ref{fig:latticemulti} (g)].
This phase is the SODSC phase \cite{PhysRevB.100.020509}, where two or four Majorana cones are on the surface in addition to a Majorana-zero-mode arc on the hinge connecting the surface nodes.
The split of surface Majorana nodes is approximately proportional to the Zeeman field normal to the surface (e.g., on a surface normal to the $y$ axis, the Majorana surface nodes split along the $k_z$ direction by $\propto\Delta_\text{Z}n_y$).
This phase is unique to 3d systems in the sense that there is no correspondence in 2d systems.
In the geometry in Fig.~\ref{fig:multilayer} (right), the Majorana hinge zero modes can be seen and they connect the surface Majorana nodes [Fig.~\ref{fig:latticemulti} (h)].
From this result, the SC phase in the effective model [the SC phase around the origin in Fig.~\ref{fig:effective_phasediagram} (a)] can be a SODSC phase.
However, we couldn't identify this phase with our technique and we leave this problem for future work.

\section{Conclusion}
\label{sec:conclusion}

In this paper, we studied a multilayer model of TI and SC layers where the proximity-induced pair potential has alternating signs at the top and bottom surface of each TI
and under in-plane Zeeman coupling.
In this model, we found three types of 3d HOTSC phases: 
(i) a SOTSC phase, each $k_z$ section of which is a 2d SOTSC and that has Majorana hinge zero modes forming flat bands in the whole $k_z$ space,
(ii) a SODSC phase, which has surface Majorana cones and a Majorana hinge arc connecting them,
and (iii) NLSC phases that have drumhead states on the surface and Majorana hinge zero modes forming arcs connecting edges of the drumhead states.
The SOTSC phase appears by applying in-plane Zeeman coupling in a 3d helical TSC phase, while NLSC phases with drumhead and hinge modes emerge from WSC phases and intervene between the SOTSC phase and NLSC phases without hinge states.
On the other hand, the SODSC phase appears around the strong and weak TI phases when both Zeeman coupling and the proximity effect are small.
The SODSC phase and the NLSC phase with drumhead and hinge modes are unique to 3d systems.

As a lattice model, we employed Bi$_2$Se$_3$-type TI for numerical study, but the theory could be applied to multilayers with other TIs.
In that case, the relative position of the Majorana hinge zero modes against the direction of the Zeeman field could be different from ours as it is dependent on the details of the electronic structure.
From a practical perspective, a multilayer of Bi$_2$Se$_3$ could be constructed with superconducting NbSe$_2$ \cite{PhysRevLett.114.017001} or Pb \cite{Trang2020}.
In addition, since
Zeeman coupling can be enhanced by doping magnetic impurities to the Bi$_2$Se$_3$ layers \cite{doi:10.1126/science.1187485,doi:10.1126/science.1189924}
and since the HOTSC phases appear in small $\Delta_\text{Z}$ regions [Fig.~\ref{fig:latticemulti} (b), (c), and (d)],
it is enough to induce Zeeman coupling which is smaller than the induced pair potential.
A small magnetic field would not disturb the proximity effect since the critical field is enhanced as SC layers are thinner,
and since the motion of the surface Dirac electron is less affected by the magnetic field applied in parallel.

A $\pi$-junction between neighboring SC layers could be realized by connecting them through an SC ring and by threading a magnetic flux (see a figure in \cite{PhysRevB.96.019901}).
When SC loops are placed vertically, the applied in-plane magnetic field generates $\pi$-phase difference and simultaneously works as an in-plane Zeeman field.
Notice that once SC loops are fabricated, the $\pi$-junction condition restricts the magnetic field strength, and hence in-plane Zeeman coupling.
Fortunately, however, fine-tuning the Zeeman coupling strength is not necessary, since any of the HOTSC phases realized in our model persists down to small $\Delta_\text{Z}$ limit [Fig.~\ref{fig:latticemulti} (b), (c), and (d)].
In addition, the presence of Majorana hinge flat modes, which is essential in higher-order topological phases, is considered to persist even if the phase difference is not exactly $\pi$ as long as the bulk energy gap of corresponding $k_z$ sections remains open (see discussion for the 2d case in \cite{PhysRevLett.122.126402}.)
Notice however that the other Majorana modes such as the bulk nodal lines and the surface drumhead states can be subject to qualitative changes by the deviation from $\pi$.

\begin{acknowledgments}
This work was supported by the Japan Society for the Promotion of Science KAKENHI (Grant Nos. JP17K17604 and JP20H01830)
and by CREST, Japan Science and Technology Agency (Grant No. JPMJCR18T2).
\end{acknowledgments}

\bibliographystyle{ieeetr}
\bibliography{reference}

\end{document}